\documentclass{article}

\usepackage{setspace}
\usepackage[english]{babel}
\usepackage[utf8]{inputenc}
\usepackage[T1]{fontenc}
\usepackage{multimedia}
\usepackage{amsfonts,multirow}
\usepackage{amsmath,bm}
\usepackage{amsthm}
\usepackage{amssymb}
\usepackage{epsfig}
\usepackage{aeguill}
\usepackage{graphicx, psfrag, color}
\usepackage{subfig}
\usepackage{color}
\usepackage{geometry}
\usepackage[superscript,biblabel]{cite}
\usepackage[super,comma]{natbib}

\vfuzz \hfuzz
\newcommand{\red}[1]{\textcolor{red}{#1}}

\renewcommand{\P}{{\rm I}\kern-0.14em{\rm P}}
\newcommand{\1}{{\rm 1}\kern-0.28em{\rm I}}
\newcommand{\E}{{\rm I}\kern-0.14em{\rm E}}
\newcommand{\R}{{\rm I}\kern-0.14em{\rm R}}
\newcommand{\reals}{{\rm I}\kern-0.16em{\rm R}}

\newcommand{\p}{{\rm I}\kern-0.18em{\rm P}}

\newcommand{\beq}{\begin{equation}}
\newcommand{\eeq}{\end{equation}}

\makeatletter
\newcommand*{\indep}{%
  \mathbin{%
    \mathpalette{\@indep}{}%
  }%
}
\newcommand*{\nindep}{%
  \mathbin{
    \mathpalette{\@indep}{\not}
  }%
}
\newcommand*{\@indep}[2]{%
  \sbox0{$#1\perp\m@th$}
  \sbox2{$#1=$}
  \sbox4{$#1\vcenter{}$}
  \rlap{\copy0}
  \dimen@=\dimexpr\ht2-\ht4-.2pt\relax
  \kern\dimen@
  {#2}%
  \kern\dimen@
  \copy0 
} 
\makeatother

\usepackage{tikz}
\usetikzlibrary{arrows,shapes}
\tikzstyle{var}=[draw,circle,thick, text width=4mm, inner sep = 3pt]
\tikzstyle{varr}=[draw,circle,thick]
\tikzstyle{varf}=[draw=none,fill=none]
\tikzstyle{varCond}=[draw,rectangle,rounded corners=3pt, fill=gray, minimum width=2em,minimum height=2em]
\tikzstyle{edge} = [draw,thick,->,>=latex]
\tikzstyle{edge2} = [draw,thick,-]
\tikzstyle{edge3} = [draw,dashed,->,>=latex]  
\tikzstyle{edged} = [draw,dashed,<->,>=latex]

\title{{\bf Which practical interventions does the $do$-operator refer to in causal inference? Illustration on the example of obesity and cancer.}}
\author{Lola Etievant$^{(1)}$ and Vivian Viallon$^{(2)}$}

\date{\begin{small}
$^{(1)}$ Univ Lyon, Université Claude Bernard Lyon 1, CNRS UMR 5208, Institut Camille Jordan, 43 boulevard du 11 novembre 1918, F-69622 Villeurbanne, France.\\
$^{(2)}$ International Agency for Research on Cancer,  Nutritional Methodology and Biostatistics Group, Lyon, France.
\end{small}
}

\begin{document}
\maketitle


\begin{abstract}
For exposures $X$ like obesity, no precise and unambiguous definition exists for the hypothetical intervention $do(X=x_0)$. This has raised concerns about the relevance of causal effects estimated from observational studies for such exposures. Under the framework of structural causal models, we study how the effect of $do(X=x_0)$ relates to the effect of interventions on causes of $X$. We show that for interventions focusing on causes of $X$ that affect the outcome through $X$ only, the effect of $do(X=x_0)$ equals the effect of the considered intervention. On the other hand, for interventions on causes $W$ of $X$ that affect the outcome not only through $X$, we show that the effect of $do(X=x_0)$ only partly captures the effect of the intervention. In particular, under simple causal models (e.g., linear models with no interaction), the effect of $do(X=x_0)$ can be seen as an indirect effect of the intervention on $W$. 
\end{abstract}

\section{Introduction}

Because most epidemiological results are derived from observational data, their causal interpretation has always been at the center of concern \cite{rothman_modern_2008}. Causal inference theory, which has attracted a lot of interest in the last few decades, has proved useful to formally describe conditions ensuring the causal validity of results derived from observational data \cite{rubin_estimating_1974, Pearl_2000, rothman2005causation, glymour_causal_2008, Pearl_2009, Hernan_Robins_Book}. For example, a number of sets of sufficient conditions has been established for the identifiability of causal effects in the presence of confounding or non-random selection. Under the so-called Structural Causal Models \cite{Pearl_2000, Pearl_2009} (SCMs), and further assuming that the structure of the underlying Directed Acyclic Graph (DAG) is known, a key condition for the identifiability of the causal effect  is exchangeability, or ignorability \cite{Pearl_2000, Pearl_2009, Hernan_Robins_Book}. In particular, exchangeability has been shown to hold conditionally on any set of variables satisfying the back-door criterion \cite{Pearl_2000, Pearl_2009}. Then, a variety of statistical approaches have been proposed for the estimation of causal effects under increasingly complex settings including time-varying confounding, failure time data, etc. Among other approaches, we shall mention the parametric g-formula, inverse probability weighting approaches, g-estimation and doubly robust procedures \cite{Pearl_2000, Hernan_Robins_Book, lunceford2004stratification}.  

Even if their use has been controversial \cite{dawid2000causal}, counterfactual variables, or potential outcomes, are key to most causal inference theories commonly considered nowadays, in epidemiology, social science, statistics and computer science. The $do$-calculus that accompanies SCMs allows precise definitions of these variables and their joint distribution \cite{Pearl_2000, Pearl_2009}. Here, we will use the notation $Y^{(X=x_0)}$ to denote the counterfactual variable representing the outcome that would have been observed in the counterfactual world $\Omega^{(X=x_0)}$ that would have followed the hypothetical intervention $do(X=x_0)$, where $X$ is the exposure of interest and $x_0$ is any potential value for this exposure \cite{Pearl_2000}. For simplicity, we will focus on binary outcomes, and we let $\p(Y=1 | do(X=x_0)=1) = \p(Y^{(X=x_0)} =1)$ denote the probability of observing the outcome in this counterfactual world.

For some exposures, the lack of a precise and unambiguous definition for the intervention $do(X=x_0)$ has raised some concerns in the literature \cite{cole2009consistency, hernan2011compound, petersen2011compound, petersen2014causal, van2005discussion, pearl2010consistency, vanderweele2013causal, Vandenbroucke, hernan2016does, Hernan_2008}.  For example, consider the case where $X$ stands for a binary variable indicating obesity status at 20 years of age. In a population of lean teenagers, or even newborns, the hypothetical intervention $do(X=x_0)$, for $x_0=0$ (or $x_0=1$), could then correspond to a typically adaptive and dynamic intervention that would ensure that individuals stay lean  (or get obese) by the age of 20. However, these interventions are not well-defined, in the sense that different ``versions'' may lead to the same obesity value $x_0$ at 20 years-old. For instance, in the  ``stay lean'' arm ($do(X=0)$), individuals may be asked to do 45 minutes of physical exercise a day, or 72 minutes of physical exercise a day. They could also be asked to adhere to a healthy diet, etc. In addition, some of the versions ensuring that $X=0$ at 20 years old may be impossible to apply in practice, such as those involving genetic factors. 

More generally, this situation of a treatment with different versions, or compound treatment,  violates the ``no-multiple-versions-of-treatment assumption'', which is part of the ``Stable Unit Treatment Value Assumption'' (SUTVA) \cite{rubin1980randomization,vanderweele2013causal}. This has led to some debate around the relevance, for public health matters, of the causal effects  estimated from observational studies  in such cases. Interestingly, most arguments have been based by considering the situation where ``treatment precedes versions of that treatment'', while situations where ``versions precede treatment'' were only quickly mentioned, if at all \cite{hernan2011compound, petersen2011compound,vanderweele2013causal}. Here, we consider the situations where versions precede treatment, in which case these versions can be seen as particular levels for the causes of $X$. Then, focusing on situations where direct interventions on $X$ are impractical, we inspect how the effect of the hypothetical intervention $do(X=x_0)$ relates to the effects of interventions on causes of $X$. We show that the effect of the hypothetical intervention $do(X=x_0)$ equals the effect of particular interventions on causes of $X$ that are causes of $Y$ through $X$ only, as expected. However, for causes $W$ that influence $Y$ not only through $X$, the causal effect of $X$ differs from the causal effect of interventions on $W$. For example, in the particular case of obesity and cancer occurence, the effect of $do(X=x_0)$ is different from the effects of interventions on diet or physical activity, except for cancers whose risk is not directly associated with diet and/or physical activity.

To make our illustrative example even more concrete, we assume throughout that we intend to estimate the causal effect of obesity at 20 years of age on the occurence of cancer by the age of 50. A typical prospective cohort study would sample individuals who are cancer-free at the age of 20, record information regarding their obesity status and other variables (potential confounders, etc.) at inclusion, follow these individuals over the age interval 20-50 and finally record cancer occurence by the age of 50. Denote by $X\in\{0,1\}$ and $Y\in \{0,1\}$ the binary variables representing obesity at 20 and cancer occurence between 20 and 50. For simplicity, we further assume the absence of competing events and censoring.  

The rest of the article is organized as follows. Even if this is highly unlikely in our illustrative example, we start by considering the unconfounded setting where all causes of $X$ are causes of $Y$ through $X$ only. Then, in Section \ref{sec:confounding}, we consider a more realistic setting where confounders are present. We shall stress that this second setting is still an over-simplified version of the causal model in our illustrative example (see the Discussion). Yet, we believe it is instructive to describe the relationship between the intervention $do(X=x_0)$ and its multiple versions. Under both settings, we consider the situation where some causes are modifiable, while others are not. Section \ref{sec:discussion} presents some concluding remarks and discussion. Proofs of our main results are presented in the Appendix. 

\section{The unconfounded case}\label{sec:noconfounding}

Because exposure $X$ is not randomized in our prospective cohort study, identifiability of the causal effect of $X$ on $Y$ is generally not guaranteed. A particular situation when this causal effect is identifiable is when all causes of $X$, denoted by $U$ in this simple case, are causes of $Y$ through $X$ only. Even if this absence of confounders is highly unlikely in our illustrative example, it is instructive to consider this simple situation as a starting point. The more general situation where confounding is present is deferred to Section \ref{sec:confounding}.

\subsection{Preliminary derivations}

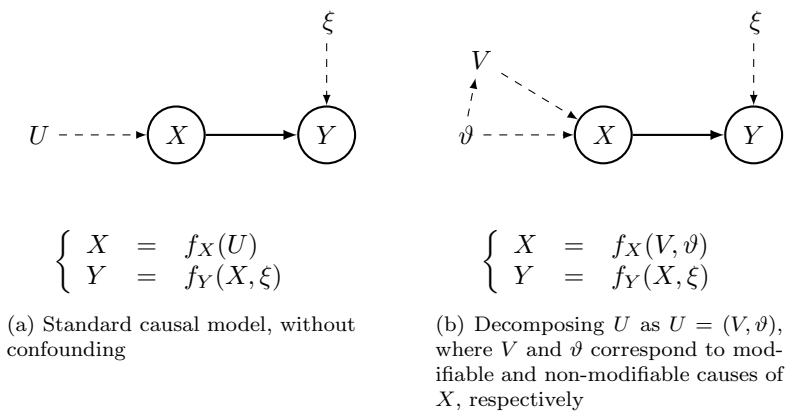
\begin{figure}
\centering
\subfloat[Standard causal model, without confounding]{\label{fig:figSimple_a}
\begin{tikzpicture}[scale=1, auto,swap, baseline]
\node[var] (X)at(0,0){$\, X$};
\node[var] (Y)at(2,0){$\, Y$};
\draw[edge] (X)--(Y);
\node[varf] (UY)at(2,1.5){$\, \xi$};
\node[varf] (UX)at(-1.85,0){$\, U$};
\draw[edge3] (UX)--(X);
\draw[edge3] (UY)--(Y);
\node[text width=4cm, align=left]at(0, -1.5){\begin{displaymath}\left\{ \begin{array}{l c l} X &=& f_X(U)\\Y &=& f_Y(X,\xi) \end{array} \right.
\end{displaymath} }; 
\end{tikzpicture}} 
\quad\quad\quad 
\subfloat[Decomposing $U$ as $U=(V, \vartheta)$, where $V$ and $\vartheta$ correspond to modifiable and non-modifiable causes of $X$, respectively]{\label{fig:figSimple_b}
\begin{tikzpicture}[scale=1, auto,swap, baseline]
\node[var] (X)at(0,0){$\, X$};
\node[var] (Y)at(2,0){$\, Y$};
\draw[edge] (X)--(Y);
\node[varf] (UY)at(2,1.5){$\, \xi$};
\node[varf] (V)at(-1.65,1){$\, V$};
\node[varf] (UX)at(-1.85,0){$\, \vartheta$};
\draw[edge3] (V)--(X);
\draw[edge3] (UX)--(X);
\draw[edge3] (UY)--(Y);
\draw[edge3] (UX)--(V);
\node[text width=4cm, align=left]at(0, -1.5){\begin{displaymath}\left\{ \begin{array}{l c l} X &=& f_X(V, \vartheta)\\Y &=& f_Y(X,\xi) \end{array} \right.
\end{displaymath} }; 
\end{tikzpicture}}
\caption{DAGs and associated structural equations in the unconfounded case. Non-circled variables ($U$, $\xi$, $\vartheta$ and $V$) correspond to  exogeneous variables \cite{Pearl_2009}, which are generally not reported in the DAG. We use dashed-arrows to connect any such exogeneous variable to any other variable.}
\label{fig:figSimple}
\end{figure}
Consider that the data available in our cohort study are generated by a causal model with associated DAG and structural equations as presented in Figure \ref{fig:figSimple_a}. Variables $\xi$ and $U$ represent all causes of $Y$ and $X$, respectively, and are assumed to be independent to each other. Both $\xi$ and $U$ may include purely random components. Given the structural equations attached to this simple causal model, we have $\{X=x\} \Rightarrow \{Y = Y^{(x)}\}$, so that consistency holds. Moreover, under this causal model, the ignorability condition $Y^{(x)}\indep X$ holds. Then, whenever the positivity condition further holds ($0<\P(X=1)<1$), we have
\begin{align*}
ACE &=\P(Y=1| do(X=1)) - \P(Y=1| do(X=0))\\ 
&= \P(Y  = 1| X=1) - \P(Y  = 1| X=0),
\end{align*}
and the causal effect of $X$ on $Y$ is identifiable. But, when direct interventions on $X$ are impractical, and only interventions on the causes of $X$ are practical, a natural question is the meaning of the hypothetical intervention $do(X=x)$. Consider the structural equation pertaining to exposure, $X = f_X(U)$, and set $f_X^{-1}(x_0) = \{u: f_X(u)=x_0\}$. Of course, we have $X=x_0 \Leftrightarrow U \in f_X^{-1}(x_0)$. As a result,  for any $u_{x_0} \in f_X^{-1}(x_0)$, $\P(Y=1 | do(U = u_{x_0}))  = \P(Y=1| do(X=x_0))$;  see Appendix \ref{App_Unconfound}. In this simple case, all interventions $do(U=u_{x_0})$ on the causes of $X$ which would yield $X=x_0$ share the same effect on $Y$: versions are irrelevant \cite{hernan2011compound, vanderweele2013causal}, and the causal effect $ \P(Y=1| do(X=x_0))$ estimated on the cohort is an estimate of this shared effect. 

\subsection{Distinguishing modifiable and non-modifiable causes}\label{sec:noconf_modvsnonmod}

To gain insight from a practical standpoint, the previous analysis can be slightly refined by decomposing causes of $X$ as $U=(V, \vartheta)$ where $V$ and $\vartheta$ correspond to sets of modifiable and non-modifiable causes of $X$, respectively.  See Figure \ref{fig:figSimple_b}. Because non-modifiable causes may affect modifiable ones, while the former are unlikely to be affected by the latter, we do not consider  the possibility of an arrow pointing from $V$ to $\vartheta$ in Figure \ref{fig:figSimple_b}. Causes $\vartheta$ are non-modifiable and the only interventions that could be practically set up are those on $V$. Denote the set of possible values of $\vartheta$ by ${\cal V}$. Then, for any $x\in\{0, 1\}$ and $\nu\in{\cal V}$, set $f_{X|\vartheta}^{-1}(x; \nu)=\{v: f_X(v,\nu)=x\}$. First assume that this set is non-empty for any $x\in\{0, 1\}$ and $\nu\in{\cal V}$: in other words, first assume that, for any $x\in\{0, 1\}$, and for any value $\nu$ for the non-modifiable factors $\vartheta$, there exists some value $v$  of the modifiable factors $V$ such that $f_X(\nu, v)=x$. Now, for individuals such that $\vartheta=\nu_0$, for any $\nu_0\in {\cal V}$, we have $X=x_0 \Leftrightarrow V  \in f_{X|\vartheta}^{-1}(x_0; \nu_0)$. Therefore $\P(Y^{(V = v_{x_0}(\nu_0))}=1 | \vartheta=\nu_0) = \P(Y=1 | do(V = v_{x_0}(\nu_0)), \vartheta=\nu_0) = \P(Y=1| do(X=x_0))$ for any $v_{x_0}(\nu_0) \in f_{X|\vartheta}^{-1}(x_0; \nu_0)$. Denote by $do(V = v_{x_0}(\vartheta))$ any intervention which sets, for all individuals in the population, the value of  $V$ according to the value  $\nu_0$ of $\vartheta$, in such a way that for any individual with $\vartheta=\nu_0$, the intervention $do(V = v_{x_0}(\vartheta))$ sets $V$ to  $v_{x_0}(\nu_0) \in f_{X|\vartheta}^{-1}(x_0; \nu_0)$. Then, we have $\P(Y=1 | do(V = v_{x_0}(\vartheta))) =  \P(Y=1| do(X=x_0))$. In other words, versions are again irrelevant and any such intervention has the same effect on $Y$, which is $ \P(Y = 1| do(V = v_{x_0}(\vartheta)) = \P(Y=1| do(X=x_0))$.

Of course, unless there exists at least one value $v_1\in \cap_{\nu\in {\cal V}} \{f_{X|\vartheta}^{-1}(x_0; \nu)\}$,  only a dynamic, {\em i.e.} individual-specific, treatment can be adopted to attain this effect. 
For instance, consider the ``stay lean'' arm of the clinical trial mentioned in the Introduction. Because individuals may be more or less genetically predisposed to obesity, some individuals will have to make little effort to stay lean by the age of 20, while others will have to adopt a drastic diet and/or have intense physical activity, etc. We may stress that this heterogeneity among individuals is at the core of  personalized (preventive) medicine and need to be acknowledged, rather than discarded, in causal inference. 
Similarly, our cohort reflects this heterogeneity: individuals sharing the same obesity status $\{X=x_0\}$, for $x_0 \in \lbrace 0,1 \rbrace$, can differ regarding $V$ and $\vartheta$. More precisely,
for $x_0\in\{0, 1\}$, set ${\cal V}(x_0)=\{\nu \in{\cal V}: f_{X|\vartheta}^{-1}(x_0; \nu) \neq \emptyset\}$. The lean and obese groups in our cohort are sampled from
\begin{align*}
\{X=x_0\} &= \bigcup_{\nu \in{\cal V}(x_0)} \left\{ \{\vartheta = \nu\} \bigcap \{V \in f_{X|\vartheta}^{-1}(x_0; \nu)\}\right\}
\end{align*}
for $x_0=0$ and $x_0=1$, respectively. Again, if the model of Figure \ref{fig:figSimple_b} is correct, versions of the compound treatment obesity are not relevant \cite{hernan2011compound, vanderweele2013causal}. Therefore, how the levels of the causes of ``obesity at 20 years of age'' are mixed up in the group of obese, or lean, individuals in our cohort is not relevant either: our cohort would return unbiased estimates for the quantity $\P(Y=1| do(X=x_0))=\P(Y=1| X=x_0)$, just as the clinical trial would. Then, the effect of the intervention $do(X=x_0)$ can again be interpreted as the effect of any intervention on the causes of $X$ ensuring $X=x_0$. 


If, for some $x$,  there exist some values $\nu_1\in {\cal V}$ of the non-modifiable variables $\vartheta$ such that the set  $f_{X|\vartheta}^{-1}(x; \nu_1)$ is empty, the intervention $do(X=x)$ is purely theoretical for individuals such that $\vartheta=\nu_1$  since no practical intervention could yield $X=x$ for them. However, under the assumptions of SCMs, and if the DAG of Figure \ref{fig:figSimple_b} is correct, the effect of the hypothetical intervention $do(X=x_0)$ can still be estimated from our cohort study even if no practical intervention ensuring $X=x_0$ exists for individuals with $\vartheta=\nu_1$. Indeed, we have $\P(Y=1| do(X=x_0), \vartheta = \nu_1) =\P(Y=1| do(X=x_0)=\P(Y=1| X=x_0)$.

\section{The more standard case with confounders}\label{sec:confounding}

\subsection{Preliminary analyses}\label{sec:Confounding0}

\begin{figure}
\centering
\subfloat[Standard causal model with confounding]{\label{fig:figConfound_a}
\begin{tikzpicture}[scale=1, auto,swap, baseline]
\node[var] (X)at(0,0){$\, X$};
\node[var] (Y)at(2,0){$\, Y$};
\node[var] (W)at(1,1.5){$\, W$};
\draw[edge] (X)--(Y);
\draw[edge] (W)--(Y);
\draw[edge] (W)--(X);
\node[varf] (UY)at(2,1.5){$\, \xi$};
\node[varf] (UX)at(-1.85,0){$\, U$};
\draw[edge3] (UX)--(X);
\draw[edge3] (UY)--(Y);
\node[text width=4cm, align=left]at(0, -1.5){\begin{displaymath}\left\{ \begin{array}{l c l} X &=& f_X(U, W)\\Y &=& f_Y(X,W,\xi) \end{array} \right.
\end{displaymath} }; 

\end{tikzpicture}} 
\quad\quad\quad 
\subfloat[Distinguishing modifiable and non-modifiable causes of $X$ in the presence of confounding]{\label{fig:figConfound_b}
\begin{tikzpicture}[scale=1, auto,swap, baseline]
\node[var] (X)at(0,0){$\, X$};
\node[var] (Y)at(2,0){$\, Y$};
\node[var] (W)at(0.35,1.5){$W$};
\node[var] (om1)at(1.75,1.5){$\, Z$};
\draw[edge] (X)--(Y);
\node[varf] (UY)at (3,1.5){$\, \xi$};
\node[varf] (V)at(-1.65,1){$\, V$};
\node[varf] (UX)at(-1.85,0){$\, \vartheta$};
\draw[edge] (W)--(Y);
\draw[edge] (W)--(X);
\draw[edge] (om1)--(Y);
\draw[edge] (om1)--(X);
\draw[edge3] (V)--(X);
\draw[edge3] (UX)--(X);
\draw[edge3] (UY)--(Y);
\draw[edge] (om1)--(W);
\draw[edge3] (UX)--(V);
\node[text width=4cm, align=left]at(0, -1.5){\begin{displaymath}\left\{ \begin{array}{l c l} X &=& f_X(V, \vartheta, W, Z)\\Y &=& f_Y(X,W,Z, \xi) \end{array} \right.
\end{displaymath} }; 
\end{tikzpicture}}
\caption{Causal models in the presence of confounders.}
\label{fig:figConfound}
\end{figure}
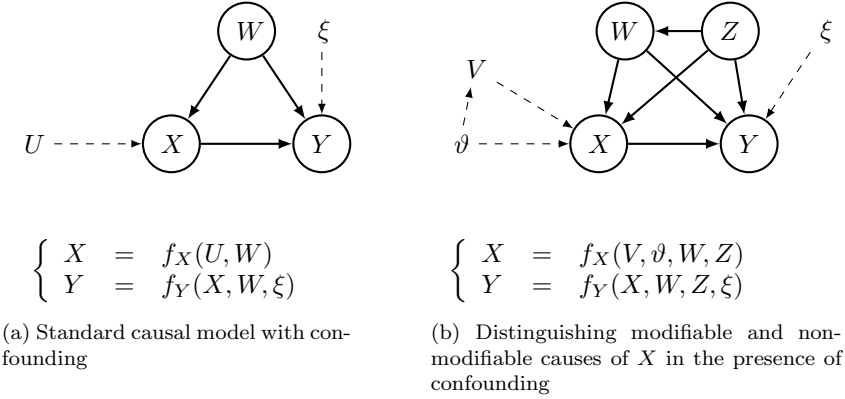

We now turn our attention to the more common situation where confounding is present. Without loss of generality, assume that causes of $X$ are grouped in two sets, $W$ and $U$. Here, and as above, causes in $U$ are assumed to have an effect on $Y$ through $X$ only, while $W$ is the set of common causes of $X$ and $Y$, that is the set of confounders in the $X$-$Y$ relationship. In our illustrative example, $W$ could include gender, physical activity and dietary habit, while $U$ might include genetic predisposition to obesity. Figure \ref{fig:figConfound_a} depicts the corresponding causal model. Assume for ease of notation that the set ${\cal W}$ of possible values for $W$ is discrete. Further recall that consistency still holds, and assume  that $0<\P(X=1|W=w)<1$ for all $w$ such that $\P(W=w)>0$. Then, because $Y^{(x)}\indep X|W$ under the model depicted in Figure \ref{fig:figConfound_a} , the causal effect of $X$ on $Y$ is identifiable. More precisely, we have
\begin{align*}
ACE 
&= \sum_w [\P(Y  = 1| X=1, W=w) - \P(Y  = 1| X=0, W=w)]\P(W=w).
\end{align*}
But, again, a natural question is how the hypothetical intervention $do(X=x)$ does relate to interventions on causes of $X$. Neglecting for now issues related to the possibility to apply these interventions in practice, these interventions can concern either $(i)$ $U$ only, $(ii)$ $W$ only, or $(iii)$ both $U$ and $W$.

First consider interventions on $U$ only and set, for any $x\in\{0, 1\}$ and $w\in{\cal W}$, $f_{X|W}^{-1}(x;w)=\{u: f_X(u,w)=x\}$. For any $w_0\in{\cal W}$, we have $X=x_0 \Leftrightarrow U  \in f_{X|W}^{-1}(x_0;w_0)$ for individuals belonging to stratum $W=w_0$. Then, assume that $f_{X|W}^{-1}(x_0;w_0)$ is non-empty for all $(x_0, w_0) \in\{0,1\}\times{\cal W}$ and denote by $do(U = u_{x_0}(W))$ any intervention setting  $U$ to any value $u_{x_0}(w_0) \in f_{X|W}^{-1}(x_0;w_0)$ for individuals in stratum $W=w_0$, for all $w_0\in {\cal W}$. Arguing as in Section \ref{sec:noconf_modvsnonmod}, we get $\P(Y=1 | do(U = u_{x_0}(W))) =  \P(Y=1| do(X=x_0))$; see Section \ref{App_Confound_type_i} in the Appendix. Again, versions are irrelevant, and any such intervention has the same effect on $Y$, which is $\P(Y=1| do(X=x_0))$. 

Now consider interventions on $W$ only and set, for any $x\in\{0, 1\}$ and $u\in{\cal U}$, $f_{X|U}^{-1}(x;u)=\{w: f_X(u,w)=x\}$. Then, assume that $f_{X|U}^{-1}(x;u)$ is non-empty for every $(x, u)\in\{0,1\}\times{\cal U}$, and for any $u_0\in {\cal U}$, denote by $w_{x_0}(u_0)$ one given element of $f_{X|U}^{-1}(x_0;u_0)$. Given this particular collection of values $(w_{x_0}(u))_{u\in{\cal U}}$, denote by $do(W = w_{x_0}(U))$ the intervention which sets  $W$ to $w_{x_0}(u_0)$ for individuals in stratum $U=u_0$, for all $u_0\in {\cal U}$. Arguing as before, it comes that $\P(Y=1 | do(W = w_{x_0}(U))) = \P(Y=1 | do(X=x_0, W=w_{x_0}(U)))$, which generally differs from $\P(Y=1 | do(X=x_0))$.  The intervention $do(W = w_{x_0}(U))$ does entail $X=x_0$ for all individuals, but because $W$ has an effect on $Y$ not only through $X$, the effect of $do(W = w_{x_0}(U))$ is not entirely captured by that of $do(X=x_0)$. Actually, $X$ can be seen as a mediator in the $W-Y$ relationship, and, under simple models, in particular in the absence of interaction between $X$ and $W$, the effect of $do(X=x_0)$ is actually related to the indirect effect of the intervention $do(W = w_{x_0}(U))$, through $X$; see Section \ref{App_Indir} in the Appendix. It is also important to note that $\P(Y=1 | do(W = w_{x_0}(U))$ depends on the collection of values $(w_{x_0}(u))_{u\in{\cal U}}$. If  $w_0$ and $\tilde w_0$ are two distinct elements of $ f_{X|U}^{-1}(x_0; u_0)$ for some $u_0\in{\cal U}$, then $\P(Y=1 | do(W = w_0), U=u_0) = \P(Y^{(W=w_0, X=x_0)}=1)$, while $\P(Y=1 | do(W = \tilde w_0), U=u_0) = \P(Y^{(W=\tilde w_0, X=x_0)}=1)$. The difference between these two quantities is related to the direct effect of $W$, and reflects the fact that two interventions on $W$ sharing the same effect on $X$ do not necessarily have the same effects on $Y$ when $W$ has a direct effect on $Y$: in this case, versions of the compound treatment are relevant. 



Now, if $f_{X|U}^{-1}(x;u)$ is  empty for some $(x, u)\in\{0,1\}\times{\cal U}$, then no intervention on $W$ only can ensure $X=x$ for individuals in stratum $U=u$. Similarly, if $f_{X|W}^{-1}(x;w)$ is empty for some pair $(x, w)$, then no intervention on $U$ only can ensure $X=x$ for individuals in stratum $W=w$. 
Then, consider interventions on both $W$ and $U$, and set $f_X^{-1}(x)=\{(w,u): f_X(u,w)=x\}$. For any $(w_0, u_0)\in f_X^{-1}(x_0)$, it is easy to show that $\P(Y=1 | do(W = w_0, U=u_0)) = \P(Y^{(W=w_0, X=x_0)}=1)$. Therefore, interventions on both $W$ and $U$ that ensure $X=x_0$ are similar to interventions on $W$ only: their effects are generally not uniquely defined (they depend on the particular pair of values $(w_0, u_0)\in f_X^{-1}(x_0)$)  and only partly capture the effect of interventions on $X$.

\subsection{Distinguishing modifiable and non-modifiables causes}
All the analyses above can be refined by acknowledging that some causes in $U$ and $W$ are modifiable, while others are not, and by considering interventions on modifiable causes only. See Figure \ref{fig:figConfound_b}. Compared to Section \ref{sec:Confounding0}, notations become a little more complex, but conclusions remain mostly similar. For instance, consider interventions on both $V$ and $W$, where $V$ is a modifiable cause of $X$ with no direct effect on $Y$, while $W$ is a modifiable confounder in the $X-Y$ relationship. For any $x_0\in\{0,1\}$ and any potential values $\nu$ and $z$ for non-modifiable causes $\vartheta$ and $Z$, assume that the set $f_{X|\vartheta, Z}^{-1}(x_0;\nu,z) = \{(v,w): f_X(v, \nu, w, z)=x_0\}$ is non-empty, and denote by $(v_{x_0}(\nu, z), w_{x_0}(\nu, z))$ one given element in this set. Then denote by $do(V=v_{x_0}(\vartheta, Z), W=w_{x_0}(\vartheta, Z))$ the intervention setting $V$ to $v_{x_0}(\nu_0, z_0)$ and $W$ to $w_{x_0}(\nu_0, z_0)$ for any individuals in stratum $\{\vartheta=\nu_0\}\cap \{Z=z_0\}$, for all $\nu_0, z_0$. Arguing as before, it can be shown that $\P(Y=1 | do(V=v_{x_0}(\vartheta, Z), W=w_{x_0}(\vartheta, Z))) = \P(Y=1 | do(X=x_0, W=w_{x_0}(\vartheta, Z))).$  This quantity generally differs from $\P(Y=1 | do(X=x_0)$ and the reason again is that the intervention $do(V=v_{x_0}(\vartheta, Z), W=w_{x_0}(\vartheta, Z))$ not only ensures that $X=x_0$, but it also has a direct effect on $Y$ through the intervention on $W$. 

\section{Conclusion-Discussion}\label{sec:discussion}

In this article, we showed how the hypothetical intervention $do(X=x_0)$, when impossible to apply in practice, relates to interventions on causes of $X$. Basing our arguments on structural causal models, our conclusions are in line with those of Petersen \cite{petersen2011compound}: the DAG which represents our assumptions on the causal model under study is basically sufficient (and necessary) to precisely understand how $do(X=x_0)$ can be interpreted. When interventions on causes of $X$ that are causes of $Y$ through $X$ only exist, the effect of $do(X=x_0)$ captures the effect of such interventions. However, for causes of $X$, say $W$, that cause $Y$ not only through $X$, the effect of $do(X=x_0)$ only partly captures the effect of interventions on $W$. Under simple causal models, the effect of $do(X=x_0)$ is related to the indirect effect of interventions on $W$.

Taking the example of obesity (at 20 years old) and the risk of cancer (by the age of 50), our results confirm concerns raised by several authors \cite{vanderweele2013causal, Hernan_2008, hernan2011compound}: because most modifiable causes of obesity can be regarded as confounders in the obesity-cancer relationship, the effect of obesity estimated from observational data likely differs from the effect of interventions on these causes, which could be estimated through clinical trials. At this point, however, we may insist on the fact that, if all modifiable causes of obesity are confounders in the obesity-cancer relationship, then clinical trials would not yield an estimate of the effect of obesity on cancer. Instead, a clinical trial would return an estimate of the causal effect of the considered intervention on cancer, and this effect would only partly capture the effect of obesity. 
Consider again the clinical trial sketched in the Introduction. More precisely, consider a randomized clinical trial where the study population, corresponding, e.g. to lean teenagers, is randomly assigned to two arms. Denote by $U$ and $Z$ the other, possibly non-modifiable, causes of $X$, with $Z$ corresponding to common causes of $Y$ and $X$, and $U$ corresponding to causes of $Y$ through $X$ only. In this setting, observe that $Y^{X=x} \nindep W$ while $Y^{X=x} \indep (W,Z)$ in general. Denote by ${\cal U}$ and ${\cal Z}$ the sets of possible values for $U$ and $Z$, respectively. Then, an ``ideal'' clinical trial would consist in randomly assigning individuals to one of the following two groups: those for whom $W$ would be set to $w_1(U, Z)$ and those for whom $W$ would be set to $w_0(U, Z)$, for two given collections of values $(w_0(u, z))_{u\in{\cal U},z\in {\cal Z}}$ and $(w_1(u, z))_{u\in{\cal U}, z\in {\cal Z}}$, where $w_0(u, z)$ and $w_1(u, z)$ ensure that $X=0$ and $X=1$, respectively, for individuals with $U=u$ and $Z=z$. Assuming complete compliance, and arguing as in Section \ref{sec:confounding}, it is easy to show that the comparison of these two groups would return an estimate of the effect of this particular intervention on $W$, 	not that of $X$. Comparisons should be made between groups of individuals sharing the same value  for $W$ and $Z$ to obtain a valid estimate of the effect of obesity, within strata defined by $W$ and $Z$. In other words, under this ideal clinical trial setting, non-modifiable confounders in the $X-Y$ relationship would still have to be measured and controlled for to unbiasedly estimate the causal effect of obesity, within strata defined by $W$ and $Z$. When controlled for a sufficient set of confounders, analyses based on observational studies can be used to derive unbiased estimates of these same effects.

There are a number of subtleties that we neglected for the sake of simplicity. First, a clinical trial whose objective is to prevent obesity by the age of 20 would typically not only be dynamic, but also adaptive, i.e. the intervention is not only subject-specific, but it is also time-dependent. A good example is the Feeding Dynamic Intervention, to prevent childhood obesity (https://clinicaltrials.gov/ct2/show/NCT01515254). Similarly, although we focused on time-fixed exposure and confounders, but they are all time-varying in the population. For instance, physical activity and food intakes vary over the age interval $[0, 20)$, and the corresponding variables are all potential confounders in the relationship between obesity at 20 years-old and cancer occurence before 50 years-old. Another important time-varying cause of obesity at 20 years-old is obesity over the age interval $[0, 19)$. Consequently, individuals in the two groups of our cohort, obese and lean at 20 years-old, do not only differ because of their status regarding obesity at 20 years of age, they also typically differ with respect to their histories regarding obesity, physical activity and dietary habits. This can lead to biases if these histories are not appropriately accounted for in the analysis \cite{etievant2018causal}. Second, selection bias may also be at play in our cohort study since only individuals who are cancer-free at 20 can be included. This selection bias will be more severe if cancer risk before 20 years old is associated to levels of obesity, physical activity and dietary habits over the age interval [0, 19]. This selection bias due to prevalent exposure and depletion of susceptibles has been put forward as one of the reasons explaining the discrepancies between results obtained through observational and interventional data when studying the association between hormone replacement therapy and coronary heart disease for instance \cite{hernan2008observational}. 

\bibliographystyle{ieeetr}
\bibliography{Viallon_etal_Intervention.bib}

\begin{thebibliography}{10}

\bibitem{rothman_modern_2008}
K.~J. Rothman, S.~Greenland, and T.~L. Lash, {\em Modern Epidemiology}.
\newblock Lippincott Williams \& Wilkins, 2008.

\bibitem{rubin_estimating_1974}
D.~B. Rubin, ``Estimating causal effects of treatments in randomized and
  nonrandomized studies,'' {\em Journal of Educational Psychology}, vol.~66,
  no.~5, pp.~688--701, 1974.

\bibitem{Pearl_2000}
J.~Pearl, {\em Causality: models, reasoning, and inference}.
\newblock Cambridge, U.K. ; New York: Cambridge University Press, 2000.

\bibitem{rothman2005causation}
K.~J. Rothman and S.~Greenland, ``Causation and causal inference in
  epidemiology,'' {\em American Journal of Public Health}, vol.~95, no.~S1,
  pp.~S144--S150, 2005.

\bibitem{glymour_causal_2008}
M.~Glymour and S.~Greenland, ``Causal diagrams,'' in {\em Modern epidemiology},
  pp.~183--209, 3rd ed. lippincott williams \& wilkins~ed., 2008.

\bibitem{Pearl_2009}
J.~Pearl, ``Causal inference in statistics: {An} overview,'' {\em Statistics
  Surveys}, vol.~3, no.~0, pp.~96--146, 2009.

\bibitem{Hernan_Robins_Book}
M.~A. Hernan and J.~M. Robins, {\em Causal Inference}.
\newblock Boca Raton: Chapman \& Hall/CRC, forthcoming.

\bibitem{lunceford2004stratification}
J.~K. Lunceford and M.~Davidian, ``Stratification and weighting via the
  propensity score in estimation of causal treatment effects: a comparative
  study,'' {\em Statistics in medicine}, vol.~23, no.~19, pp.~2937--2960, 2004.

\bibitem{dawid2000causal}
A.~P. Dawid, ``Causal inference without counterfactuals,'' {\em Journal of the
  American Statistical Association}, vol.~95, no.~450, pp.~407--424, 2000.

\bibitem{cole2009consistency}
S.~R. Cole and C.~E. Frangakis, ``The consistency statement in causal
  inference: a definition or an assumption?,'' {\em Epidemiology}, vol.~20,
  no.~1, pp.~3--5, 2009.

\bibitem{hernan2011compound}
M.~A. Hernan and T.~J. VanderWeele, ``Compound treatments and transportability
  of causal inference,'' {\em Epidemiology (Cambridge, Mass.)}, vol.~22, no.~3,
  p.~368, 2011.

\bibitem{petersen2011compound}
M.~L. Petersen, ``Compound treatments, transportability, and the structural
  causal model: the power and simplicity of causal graphs,'' {\em
  Epidemiology}, vol.~22, no.~3, pp.~378--381, 2011.

\bibitem{petersen2014causal}
M.~L. Petersen and M.~J. van~der Laan, ``Causal models and learning from data:
  integrating causal modeling and statistical estimation,'' {\em Epidemiology
  (Cambridge, Mass.)}, vol.~25, no.~3, p.~418, 2014.

\bibitem{van2005discussion}
M.~van~der Laan, T.~Haight, and I.~Tager, ``Discussion: Hypothetical
  interventions to define causal effects: afterthought or prerequisite,'' {\em
  The American Journal of Epidemiology}, vol.~162, pp.~382--88, 2005.

\bibitem{pearl2010consistency}
J.~Pearl, ``On the consistency rule in causal inference: axiom, definition,
  assumption, or theorem?,'' {\em Epidemiology}, vol.~21, no.~6, pp.~872--875,
  2010.

\bibitem{vanderweele2013causal}
T.~J. VanderWeele and M.~A. Hernan, ``Causal inference under multiple versions
  of treatment,'' {\em Journal of causal inference}, vol.~1, no.~1, pp.~1--20,
  2013.

\bibitem{Vandenbroucke}
J.~P. Vandenbroucke, A.~Broadbent, and N.~Pearce, ``Causality and causal
  inference in epidemiology: the need for a pluralistic approach,'' {\em
  International Journal of Epidemiology}, vol.~45, no.~6, pp.~1776--1786, 2016.

\bibitem{hernan2016does}
M.~A. Hern{\'a}n, ``Does water kill? a call for less casual causal
  inferences,'' {\em Annals of Epidemiology}, vol.~26, no.~10, pp.~674--680,
  2016.

\bibitem{Hernan_2008}
M.~A. Hern{\'a}n and S.~L. Taubman, ``Does obesity shorten life? {The}
  importance of well-defined interventions to answer causal questions,'' {\em
  International Journal of Obesity}, vol.~32, pp.~S8--S14, 2008.

\bibitem{rubin1980randomization}
D.~B. Rubin, ``Comment on: ``randomization analysis of experimental data: The
  fisher randomization test '' by {D}. {B}asu,'' {\em Journal of the American
  Statistical Association}, vol.~75, no.~371, pp.~591--593, 1980.

\bibitem{etievant2018causal}
L.~Etievant and V.~Viallon, ``Causal inference under over-simplified
  longitudinal causal models,'' {\em arXiv preprint arXiv:1810.01294}, 2018.

\bibitem{hernan2008observational}
M.~A. Hern{\'a}n, A.~Alonso, R.~Logan, F.~Grodstein, K.~B. Michels, M.~J.
  Stampfer, W.~C. Willett, J.~E. Manson, and J.~M. Robins, ``Observational
  studies analyzed like randomized experiments: an application to
  postmenopausal hormone therapy and coronary heart disease,'' {\em
  Epidemiology (Cambridge, Mass.)}, vol.~19, no.~6, p.~766, 2008.

\end{thebibliography}

\vskip15pt
\appendix  {\noindent\Large{\bf Appendices}}
\section{Proof in the unconfounded case}\label{App_Unconfound} 
Under the model depicted in Figure \ref{fig:figSimple_a}, we have 
\begin{align*}
\P(Y=1 | do(U = u_{x_0})) &= \P(Y^{(U = u_{x_0})} =1 )\\
&=  \P(f_Y(X^{(U= u_{x_0})}, \xi) =1)\\
&=  \P(f_Y(x_0, \xi) =1)\\
&= \P(Y^{(x_0)}=1)\\
&= \P(Y=1| do(X=x_0)).
\end{align*}

\section{Proof in the confounded case} \label{App_Confound} 
\subsection{Interventions of type $(i)$} \label{App_Confound_type_i}
Assume that $f_{X|W}^{-1}(x_0; w_0)$ is non-empty for any $x_0, w_0$. Then, under the model depicted in Figure \ref{fig:figConfound_a}, we have, for any $u_{x_0}(w_0) \in f_{X|W}^{-1}(x_0;w_0)$
\begin{align*}
\P(Y=1 | &do(U= u_{x_0}(w_0)) , W=w_0) = \P(Y^{(U = u_{x_0}(w_0))} =1 | W=w_0 )\\
&=  \P( f_Y(X^{(U = u_{x_0}(w_0))}, W, \xi) =1| W=w_0)\\
&=  \P( f_Y(x_0, w_0, \xi) =1)\\
&= \P(Y^{(X=x_0, W=w_0)}=1)\\
&= \P(Y=1| do(X=x_0 , W=w_0))\\
&= \P(Y=1| do(X=x_0) , W=w_0),
\end{align*}
where the last equality follows from rule 2 of the do-calculus\cite{Pearl_2000}.

Moreover, 
\begin{align*}
\P(Y=1 | do(U =u_{x_0}(W))) &= \sum_{w_0} \P(Y=1 | do(U = u_{x_0}(w_0)), W=w_0)\P(W=w_0) \\
&= \sum_{w_0}\P(Y=1| do(X=x_0), W=w_0) \P(W=w_0)  \\
&= \P(Y=1| do(X=x_0)).
\end{align*}

\subsection{Interventions of type $(ii)$}
Assume that $f_{X|U}^{-1}(x_0; u_0)$ is non-empty for any $x_0,u_0$. Then, under the model depicted in Figure \ref{fig:figConfound_a}, we have, for any $w_{x_0}(u_0) \in f_{X|U}^{-1}(x_0;u_0)$
\begin{align*}
\P(Y=1 | &do(W = w_{x_0}(u_0)) , U=u_0) = \P(Y^{(W = w_{x_0}(u_0))} =1 | U=u_0 )\\
&=  \P(f_Y(X^{(W = w_{x_0}(u_0))}, w_{x_0}(u_0), \xi) =1 | U = u_0)\\
&=  \P(f_Y(x_0, w_{x_0}(u_0), \xi) =1 | U = u_0)\\
&=  \P(f_Y(x_0, w_{x_0}(u_0), \xi) =1)\\
&= \P(Y^{(X=x_0, W=w_{x_0}(u_0))}=1).
\end{align*}

\subsection{Relationship with indirect effects} \label{App_Indir}

Denote by  $(w_{1}(u_0), w_{0}(u_0))_{u_0\in{\cal U}}$ two given collection of values such that $w_{1}(u_0) \in f_{X|U}^{-1}(1; u_0)$ and $w_{0}(u_0) \in f_{X|U}^{-1}(0; u_0)$. Further let $do(W=w_1(U))$ and $do(W=w_0(U))$ denote two given interventions setting  $W$ to  $w_{1}(u_0) \in f_{X|U}^{-1}(1; u_0)$ and $w_{0}(u_0) \in f_{X|U}^{-1}(0; u_0)$, respectively, for individuals in stratum $U=u_0$, for all $u_0\in {\cal U}$. We have
\begin{align*}
\E(Y^{(w_1(U))} - Y^{(w_0(U))}) &= \sum_{u}  \E(Y^{(w_1(u))} - Y^{(w_0(u)) }|U=u)\P(U=u)  \\
&=\sum_{u}  \E(Y^{(w_1(u), X^{(w_1(u))})} -  Y^{(w_0(u), X^{(w_0(u))})} |U=u)\P(U=u) \\
&=\sum_{u}  \{ \E(Y^{(w_1(u), X^{(w_1(u))})} -  Y^{(w_1(u), X^{(w_0(u))})} |U=u) \\
&\quad\quad + \E(Y^{(w_1(u), X^{(w_0(u))})} -  Y^{(w_0(u), X^{(w_0(u))})} |U=u)\} \P(U=u) \\
&=\sum_{u}  \E(Y^{(w_1(u), x_1)} - Y^{(w_1(u), x_0)} + Y^{(w_1(u), x_0)}- Y^{(w_0(u), x_0)})\P(U=u).
\end{align*}

The term $\sum_{u}  \E(Y^{(w_1(u), x_1)} - Y^{(w_1(u), x_0)})\P(U=u)$ can be regarded as an indirect effect since the level of $W$ is held fixed and only the value of $X$ changes from $x_0$  to $x_1$ which, for individuals in stratum $U=u$, equal   $X^{(W=w_0(u))}$ and $X^{(W=w_1(u))}$ respectively. More precisely, we have
\begin{align*}
&\sum_{u} \E(Y^{(w_1(u), x_1)} - Y^{(w_1(u), x_0)}) \P(U=u)\nonumber\\
&\quad\quad= \sum_{u}\{\E(Y| W=w_1(u), X=x_1) -  \E(Y| W=w_1(u), X=x_0)\} \P(U=u). 
\end{align*}
Under the model depicted in Figure \ref{fig:figConfound_a}, recall we have
\begin{align*}
&\E(Y| do(X=x_1)) - \E(Y| do(X=x_0)) \\
&\quad\quad= \sum_w \{\E(Y| W=w, X=x_1) - \E(Y| W=w, X=x_0)\}\P(W=w). 
\end{align*}
Under simple causal models, for instance when $f_Y(W, X, \xi) = \alpha^T W + \beta X + \xi$, the two quantities, $\sum_{u} \E(Y^{(w_1(u), x_1)} - Y^{(w_1(u), x_0)}) \P(U=u)$ and $\E(Y| do(X=x_1)) - \E(Y| do(X=x_0))$, coincide and equal $\beta$. However, under more complex models, these two quantities are typically different. Even under linear models, if interaction terms of the form $\gamma^T W X$ are present  in function $f_Y$, these two terms are typically different and $\sum_{u}  \E(Y^{(w_1(u), x_1)} - Y^{(w_1(u), x_0)})\P(U=u)$ would actually depend on the collection of values $\{w_1(u), u\in{\cal U}\}$.

\end{document}